\documentclass{article}
\usepackage{arxiv}
\usepackage[utf8]{inputenc} 
\usepackage[T1]{fontenc}    
\usepackage{hyperref}       
\usepackage{url}            
\usepackage{booktabs}       
\usepackage{amsfonts}       
\usepackage{nicefrac}       
\usepackage{microtype}      
\usepackage{lipsum}         
\usepackage{graphicx}
\usepackage{doi}
\usepackage{amsmath, amssymb, amsthm}
\usepackage{bm}

\newcommand{\wref}[3]{\bibitem{#1} #2 #3}

\title{ON THE PROBLEM OF SIMPLE SHEAR OF AN INCOMPRESSIBLE VISCOELASTIC SOLID \\ UNDER FINITE DEFORMATIONS}

\author{ \href{https://orcid.org/0000-0002-8296-0907}{\includegraphics[scale=0.06]{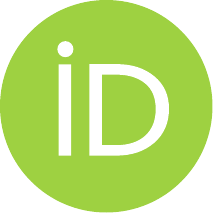}\hspace{1mm}Vladislav V.~Kozhukhov} 
        \\
	Department of Elasticity Theory\\
	Faculty of Mechanics and Mathematics\\
    Lomonosov Moscow State University\\
	Moscow, Russia 119991 \\
	\texttt{vladislav.kozhukhov@student.msu.ru} \\
}

\renewcommand{\shorttitle}{On the problem of simple shear of an incompressible viscoelastic solid under finite deformations}

\hypersetup{
pdftitle={A template for the arxiv style},
pdfsubject={q-bio.NC, q-bio.QM},
pdfauthor={C S.~Hippocampus, Elias D.~Striatum},
pdfkeywords={First keyword, Second keyword, More},
}

\begin{document}
\maketitle

\begin{abstract}
In the framework of a viscoelastic material model, whose constitutive relation is given by a one-parameter family of Gordon–Schowalter derivatives, the problem of simple shear under acceleration and random velocity motion is considered. For motion with acceleration, the presence of non-zero normal stresses is discovered, which corresponds to the Poynting effect previously discovered for this material. A problem in which the shear rate was determined as a linear function of a random variable given from a normal distribution was studied. Within the framework of the methodology proposed by V.A. Lomakin, an analytical solution of the problem is constructed. A significant dependence of the dispersion of the stress tensor components on the choice of the objective derivative was found.\end{abstract}


\section{Introduction}
This work continues the investigation of the properties of viscoelastic materials under finite deformations using a one-parameter family of Gordon–Schowalter objective derivatives [1].

To describe the behavior of viscoelastic materials under finite deformations, an approach based on the generalization of the one-dimensional elementary Maxwell model is used. Within this approach, the stress \(\sigma\) and strain \(\varepsilon\) are related by:
\begin{equation}
\dot{\sigma} = E \dot{\varepsilon} - \frac{\sigma}{T},
\end{equation}
where \(E\) is the elastic modulus, and \(T\) is the material relaxation time. For the three-dimensional case, this relation is generalized by replacing the stress \(\sigma\) with the Cauchy true stress tensor \(\mathbf{S}\), and the strain rate \(\dot{\varepsilon}\) with the strain rate tensor \(\mathbf{V}\). An objective derivative is also introduced, which accounts for the rotation and deformation of the material. In [1], the Gordon–Schowalter derivative is used as the objective derivative, written as:
\begin{equation}
D_a[\mathbf{S}] = \dot{\mathbf{S}} - \mathbf{\Omega} \mathbf{S} + \mathbf{S} \mathbf{\Omega} - a (\mathbf{V} \mathbf{S} + \mathbf{S} \mathbf{V}),
\end{equation}
where \(\mathbf{\Omega}\) is the vorticity tensor, \(\mathbf{V}\) is the strain rate tensor, and \(a\) is a scalar parameter defining the specific type of objective derivative. Depending on the value of \(a\), this derivative includes the Oldroyd, Cotter–Rivlin, and Jaumann derivatives as special cases.

Thus, the constitutive relation for viscoelastic materials under finite deformations is written as:
\begin{equation}
D_a[\mathbf{S}] = E \mathbf{V} - T^{-1} \mathbf{S},
\end{equation}
or, expanded:
\begin{equation}
\dot{\mathbf{S}} = -\mathbf{\Omega} \mathbf{S} + \mathbf{S} \mathbf{\Omega} - a (\mathbf{V} \mathbf{S} + \mathbf{S} \mathbf{V}) + E \mathbf{V} - T^{-1} \mathbf{S}.
\end{equation}

Considering the definition of the tensor \(\mathbf{J}\) as:
\begin{equation}
\mathbf{J} = \mathbf{\Omega} - a\mathbf{V},
\end{equation}
the relation between the derivative of the stress tensor and the tensor \(\mathbf{J}\) can be written:
\begin{equation}
\dot{\mathbf{S}} = -\mathbf{J}^{\mathrm{T}} \mathbf{S} + \mathbf{S} \mathbf{J} + E \mathbf{V} - T^{-1} \mathbf{S}.
\end{equation}

Hereafter, the viscoelastic material is assumed to be incompressible. Then the Cauchy stress tensor is determined up to an indeterminate spherical tensor $-p \mathbf{I}$ and can be represented as $\bm{\sigma}= \mathbf{S} - p \mathbf{I}$, where $\mathbf{S}$ is found from equation (6).
\section{Formulation of the Simple Shear Problem}
The motion law of the medium under simple shear along the \(x_2\)-axis in the \(0x_2x_3\) plane is written as:
\begin{equation}
\begin{aligned}
 \begin{cases}
{x}_{1} = \mathrm{x}_{1}, \\
{x}_{2} = \mathrm{x}_{2} + k(t) \mathrm{x}_{3}, \\
{x}_{3} = \mathrm{x}_{3},
 \end{cases}
\end{aligned}
\end{equation}
where \(\mathrm{x_i}\), \({x_i}\) are the Cartesian coordinates of a material point in the reference and current configurations, respectively.

The deformation affinor, tensors $\mathbf{V}$ and $\mathbf{J}$ take the following form:
\begin{equation}
\mathbf{A} = \begin{pmatrix}
1 & 0 & 0 \\
0 & 1 & k(t) \\
0 & 0 & 1
\end{pmatrix}, \quad
\mathbf{V} = \frac{1}{2} \dot{k}(t) \begin{pmatrix}
0 & 0 & 0 \\
0 & 0 & 1 \\
0 & 1 & 0
\end{pmatrix}, \quad \mathbf{J} = \mathbf{\Omega} - a\mathbf{V} = -\frac{1}{2} \dot{k}(t) \begin{pmatrix}
0 & 0 & 0 \\
0 & 0 & a-1 \\
0 & a+1 & 0
\end{pmatrix}.
\end{equation}
\section{Solution of the Problem for Accelerated Motion}
We assume that the motion of the medium occurs with acceleration, i.e., \(k(t)=\frac{u t^2}{2},\) where \(u=const\). Substituting (8) into (6), we arrive at the following system of differential equations for the components of the stress tensor \(\mathbf{S}(t)\):
\begin{equation}
\begin{cases}
\dot{S}_{11} = -\frac{S_{11}}{T}, \\
\dot{S}_{23} = -\frac{S_{23}}{T} + \frac{u t}{2} \left[(a+1) S_{33} - (1-a) S_{22}\right] + \frac{Eu t}{2}, \\
\dot{S}_{22} = -\frac{S_{22}}{T} + u t(a+1) S_{23}, \\
\dot{S}_{33} = -\frac{S_{33}}{T} - u t(1-a) S_{23}, \\
\dot{S}_{12} = -\frac{S_{12}}{T} + \frac{u t}{2}(1+a) S_{13}, \\
\dot{S}_{13} = -\frac{S_{13}}{T} - \frac{u t}{2}(1-a) S_{12}.
\end{cases}
\end{equation}

Consider the numerical solution of problem (9) using the SciPy library [2], assuming the initial stress tensor \(\mathbf{S}(0)=0\). Set the parameter values of model (6) as follows: \(T = 0.3 \text{(s)},\)  \(E = 10^9\) (Pa). Also, set the deterministic velocity \(u = 1 \left(\frac{\text{m}}{\text{s}}\right).\)

Figure 1 shows the change in the components of the stress tensor \(\mathbf{S}\) over time t based on numerical calculations. The following notations are adopted: values of $S_{22}(t)$ are shown with a dash-dotted line (--), values of $S_{23}(t)$ are shown with a dotted-dashed line (-.), and values of $S_{33}(t)$ are shown with a dotted line (:). Each line is colored according to the parameter $a$ (darker lines correspond to $a=1$, and lighter gray lines correspond to $a=-1$).

\begin{center}
\includegraphics[width=13cm]{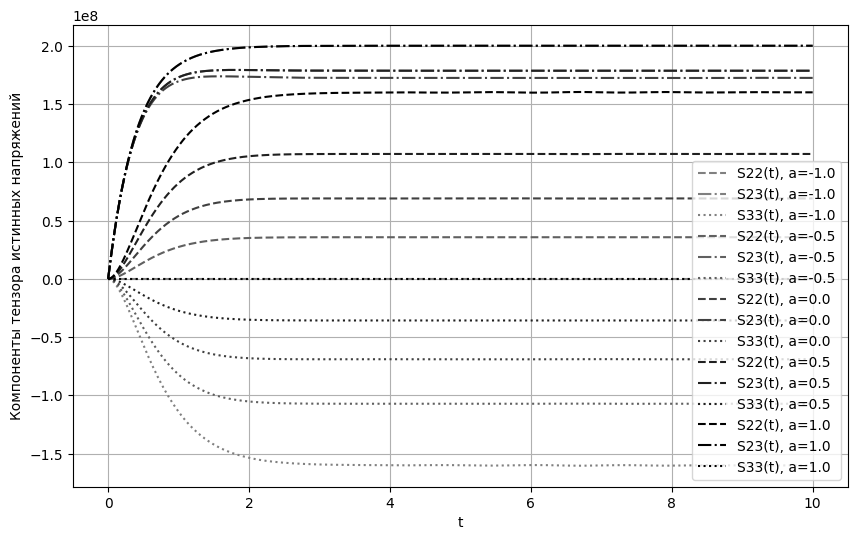}
\par Figure 1. Dependence of the components of the true stress tensor \(\mathbf{S}\) on time.
\end{center}

In Fig. 1, it can be seen that some components of the tensor \(\mathbf{S}\) are zero for all values of the parameter \(a\), specifically \(S_{11}(t)=S_{12}(t)=S_{13}(t)=0.\) For \(a = 1\), the component \(S_{33}(t)=0\); for \(a=-1\), the component \(S_{22}(t)=0\); in other cases, components \(S_{22}(t),S_{33}(t)\) have non-zero values. Moreover, note that \(S_{22}(t)=(-\frac{a+1}{1-a})S_{33}(t).\)

Thus, in the case where \(k(t)=\frac{u t^2}{2},\) with \(u=const\), for an incompressible viscoelastic material whose mechanical behavior is defined by constitutive relation (6), within the simple shear problem (7), non-zero normal stresses are observed. This corresponds to the Poynting effect for elastic media [3], previously discovered in [1].

The analytical solution to the problem was found programmatically using the SymPy library [4], which provides tools for symbolic computations. This allowed deriving the exact solution of the equations in algebraic form without resorting to numerical methods. System (9), considering the features discovered in the numerical solution, reduces to solving the following problem:

\begin{equation}
 \begin{cases}
    \dot{S}_{23}(t) = -\frac{S_{23}(t)}{T} + \frac{1}{2}  ut \left(E + 2 (1 + a) S_{33}(t) \right),\\
    \dot{S}_{33}(t) = -\frac{S_{33}(t)}{T} - u t (1 - a) S_{23}(t).
     \end{cases}
\end{equation}

The found analytical solution of system (10) is cumbersome and expressed through special functions. The constants $C_1$ and $C_2$, obtained after integration, are determined from the substitution $S_{23}(0)=0,$ $S_{33}(0)=0.$ The graphs of the components of the true stress tensor versus time coincide with the numerical solution; however, the asymptotic behavior of this analytical solution could not be established. The unknown spherical tensor $-p \mathbf{I}$ is found from the condition $\sigma_{11}(t) \equiv 0,$ hence $\bm{\sigma}(t)= \mathbf{S}(t)$.

Thus, to realize motion (7), it is necessary to specify boundary conditions in terms of stresses from solution (9) on the corresponding faces of the considered incompressible viscoelastic body. For the Cotter–Rivlin derivative ($a=-1$), we specify non-zero stresses on surfaces with normal along the $x_2$-axis in the $x_3$ direction and on surfaces with normal along the $x_3$-axis in the $x_3$ direction (with values $S_{23}, S_{33},$ respectively). For the Jaumann derivative ($a=0$), boundary conditions are specified on surfaces with normal $x_2$ in the $x_2$ direction, on surfaces with normal $x_2$ in the $x_3$ direction, and on surfaces with normal $x_3$ in the $x_3$ direction (with values $S_{22}, S_{23}, S_{33},$ respectively). For the Oldroyd derivative ($a=1$), boundary conditions are specified on surfaces with normal $x_2$ in the $x_3$ direction and on surfaces with normal $x_3$ in the $x_3$ direction (with values $S_{23}, S_{33},$ respectively).
\section{Solution of the Problem for Motion with Random Velocity from a Normal Distribution}
Problems in the mechanics of deformable bodies under random external actions are considered in the monograph by V.A. Lomakin [5]. In particular, he proved a theorem: for a linear problem in displacements within elasticity theory, when the loads applied to the body are linear functions of random variables, the solution to the problem involves finding the mean value of the stress tensor and its principal moments. In the case where the load is given as a linear function of a normally distributed random variable, it is sufficient to consider the mean value of the stress tensor and its second moment (variance of the stress tensor). In [5], it is noted that some viscoelasticity problems can be solved by this method.

We assume that \(k(t)=\lambda u t,\) where \(u=const\), \(\lambda\) is a dimensionless normally distributed random variable. Without loss of generality, let $\lambda \sim \mathcal{N}(1, 1),$ i.e., the mathematical expectation and variance of the random variable are equal to one. In this case, the mean value of the stress tensor $\mathbf{S}^{0}(t)$ is found at $u=u_0$ (the subscript $0$ is omitted for brevity):

\begin{equation}
 \begin{cases}
\dot{S}_{11} = -\frac{S_{11}}{T}, \\
\dot{S}_{23} = -\frac{S_{23}}{T} + u \left[(a+1) S_{33} - (1-a) S_{22}\right] + \frac{E  u}{2}, \\
\dot{S}_{22} = -\frac{S_{22}}{T} + u (a+1) S_{23}, \\
\dot{S}_{33} = -\frac{S_{33}}{T} -  u(1-a) S_{23}, \\
\dot{S}_{12} = -\frac{S_{12}}{T} +  u(1+a) S_{13}, \\
\dot{S}_{13} = -\frac{S_{13}}{T} -  u(1-a) S_{12}.
\end{cases}
\end{equation}

The solution to system (11) is written analytically:

\begin{equation}
\begin{aligned}
S_{22}(t) &= \frac{E e^{-\frac{(c+1) t}{T}} \left(\left(a^2-1\right) T^2 u^2 \left(e^{\frac{2 c t}{T}}-2 e^{\frac{(c+1) t}{T}}+1\right)+c \left(e^{\frac{2 c t}{T}}-1\right)\right)}{4 (a-1) (c-1) (c+1)}, \\
S_{23}(t) &= \frac{E T u e^{-\frac{(c+1) t}{T}} \left(c \left(e^{\frac{2 c t}{T}}-1\right)+e^{\frac{2 c t}{T}}-2 e^{\frac{(c+1) t}{T}}+1\right)}{4 \left(c^2-1\right)}, \\
S_{22}(t) &= -\frac{a+1}{1-a} S_{33}(t), \quad
S_{11}(t) = S_{12}(t) = S_{13}(t) = 0.
\end{aligned}
\end{equation}

where \( c = \sqrt{\left(a^2-1\right) T^2 u^2} \). This analytical solution is a variant of the solution form found in [1]. The unknown spherical tensor $-p \mathbf{I}$ is found from the condition $\sigma_{11}(t) \equiv 0,$ hence $\bm{\sigma}= \mathbf{S}$.

Since the stress tensor components (hereafter denoted $g(\lambda)$) depend on $\lambda$ in a complex manner, the condition $g(\mathbb{E}[\lambda]) = \mathbb{E}[g(\lambda)]$ generally does not hold (V.A. Lomakin's theorem for linear problems). Expand the stresses (12) in a Taylor series up to $o((u-1)^2)$ at the point $u=u_0$, and note that in this case the stresses depend only on the first power of $u$; then the condition $g(\mathbb{E}[\lambda]) = \mathbb{E}[g(\lambda)]$ is satisfied. By Jensen's inequality [6], and considering that in this problem the stress tensor components are convex upward functions (second derivative strictly less than zero), the estimate $g(\mathbb{E}[\lambda]) \geq \mathbb{E}[g(\lambda)]$ holds. Thus, an upper bound for the stress tensor components is obtained.

For the approximate calculation of the variance of a complex function \(g(\lambda)\) of a random variable \(\lambda\), the linearization method can be used:

\[
\text{Var}(g(\lambda)) \approx \left( g'(\mathbb{E}[\lambda]) \right)^2 \cdot \text{Var}(\lambda)
\]
where
 \(g'(\mathbb{E}[\lambda])\) is the derivative of the function \(g(\lambda)\) with respect to \(\lambda\), evaluated at the mathematical expectation \(\mathbb{E}[\lambda]\),
    \(\text{Var}(\lambda)\) is the variance of the random variable \(\lambda\).
    
In this problem, the dimensionless random variable $\lambda$ is independent of time $t$; however, the variance, as a function of the random variable and the deterministic solution to the problem, depends on time. We plot the variance of the stresses (normalized to the corresponding stresses) for each component of the stress tensor $\mathbf{S}(t)$ (Fig. 2).

From a practical standpoint, it is important for the stress variance to be as small as possible. Note that if $a$ tends to minus one (Cotter–Rivlin derivative), the variance of component $S_{22}$ decreases to zero, since the stress $S_{22}= 0$ at $a=-1$, while $S_{23}$ is non-zero and its variance tends to zero, but the variance of component $S_{33}$ becomes the largest. For the Oldroyd derivative ($a=1$), components $S_{22}$ and $S_{23}$ have the highest variance, while $S_{33}=0$ (with zero variance). For the Jaumann derivative ($a=0$), all stress tensor components $S_{22}$,$S_{23}$,$S_{33}$ are non-zero but have non-zero variance. Thus, there is no universal derivative from the perspective of choosing the smallest variance for all components of the stress tensor \(\mathbf{S}\). However, the specific features of each derivative may be important in practice.

\begin{center}
\includegraphics[width=17cm]{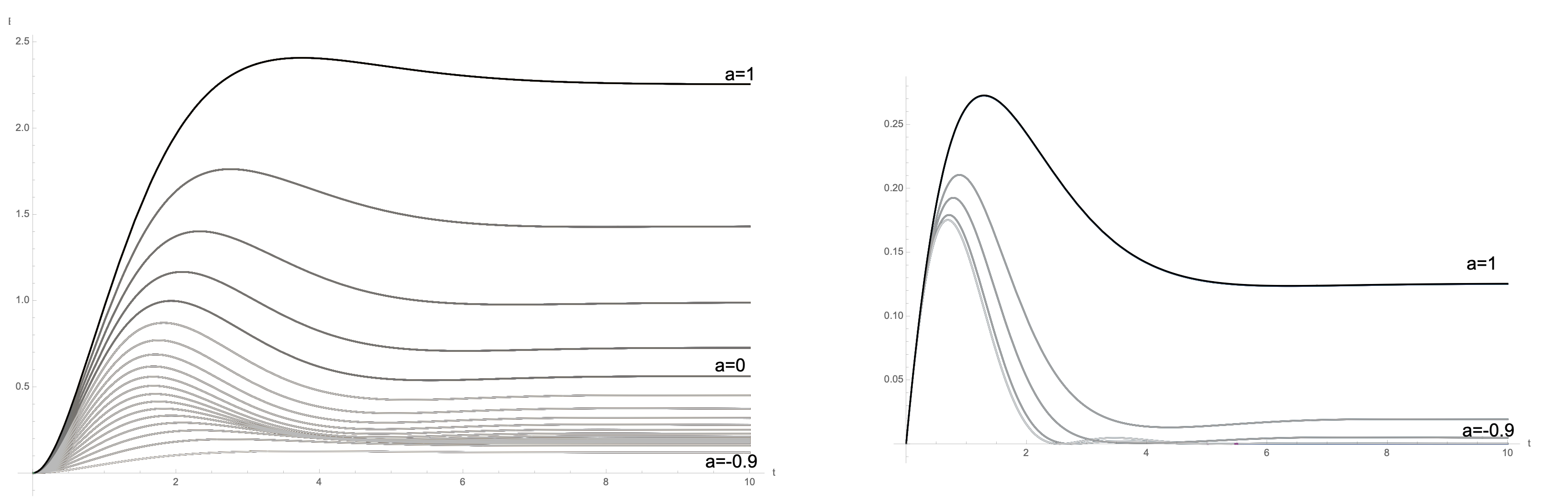}
\par Figure 2. Dependence of the dimensionless components of the squared variance of the true stress tensor \(\mathbf{S}\) on time (left graph: components $S_{22}$, right graph: $S_{23}$ as a function of parameter $a$).
\end{center}
\section{Conclusion}
Within the framework of the medium model proposed in [1], the problem of simple shear under accelerated motion is considered. The presence of non-zero normal stresses is discovered, which corresponds to the Poynting effect previously identified for this material in [1]. A problem is studied in which the shear rate is defined as a linear function of a normally distributed random variable and a constant velocity. Using the methodology proposed in [5], an analytical solution to the problem is constructed. A significant dependence of the variance of the stress tensor components \(\mathbf{S}\) on the choice of the objective derivative is revealed.

\bibliographystyle{unsrtnat}





\end{document}